\documentclass[prd,a4paper,draft,amsfonts,amssymb,amsmath,nofootinbib]{revtex4}

\newcommand{\s}{\scriptscriptstyle}
\newcommand{\kk}{/\kern-.5em k}
\newcommand{\dirac}{/\kern-.5em\partial}
\newcommand{\dd}{{\it d\kern-.35em I}}
\newcommand{\w}{\wedge \kern-.8em \wedge}
\newcommand{\D}{{\it D}}

\begin{document}

\title{Higher Derivative Fermionic Field Theories}

\author{Eduardo J. \surname{S. Villase\~nor}}
%\surname{Villase\~nor}
\email[]{eduardo@imaff.cfmac.csic.es} \affiliation{Instituto de
Matem\'aticas y F\'{\i}sica Fundamental, C.S.I.C., C/ Serrano
113bis, 28006 Madrid, Spain}

\date{May 20, 2002}

\begin{abstract}
We carry out the extension of the covariant Ostrogradski method to
fermionic field theories. Higher-derivative Lagrangians reduce to
second order differential ones with one explicit independent field
for each degree of freedom.
\end{abstract}



\maketitle

\setcounter{page}{1}

%%%%%%%%%%%%%%%%%%%%%%%%%%%%%%%%%%%%%%%%%%%%%%%%%%%%%%%%%%%%%%%%%%%%%%%%%%%%%%%%%%%%%%%%%%%%%%%%%%%%%%%%%%%%%%%%%%%%%%%%%
%%%                                                  INTRODUCTION                                                     %%%
%%%%%%%%%%%%%%%%%%%%%%%%%%%%%%%%%%%%%%%%%%%%%%%%%%%%%%%%%%%%%%%%%%%%%%%%%%%%%%%%%%%%%%%%%%%%%%%%%%%%%%%%%%%%%%%%%%%%%%%%%

\section{Introduction}

Higher derivative (HD) field theories appear in many physical
situations such as the Higgs model regularizations \cite{Higgs} or
generalized electrodynamics \cite{ED1}-\cite{julve2}, but their
main interest resides in their use as gravitational actions. HD
gravitational field models arise as effective low energy theories
of the string \cite{Gross} or are induced by quantum fields in a
curved background \cite{Birrel}. Theories with quadratic curvature
terms have been studied closely because they are renormalizable
\cite{Stelle} in four dimensions. This property led to
Renormalization-Group analyses in \cite{julve0}, which culminate
in \cite{Avr}, including attempts to avoid the appearance of Weyl
ghosts usually occurring in HD theories \cite{WG}. For these
reasons, it is in the gravitational framework where essential
progress has taken place. In particular, during the eighties a
mechanism has been devised to deal with the Hamiltonian
formulation of an arbitrary HD gauge field theory which was
successfully applied to HD gravity \cite{Lyahovich}. Somewhat
later, in the nineties, a covariant differential order reduction
for HD field theories was found \cite{ferraris}. This kind of
covariant techniques were used in the late nineties to identify
the propagating degrees of freedom (DOF) in both diff-invariant
\cite{julve1} and gauge fixed HD gravity \cite{yo1}. Moreover,
some effort has been devoted to study general bosonic HD free
theories \cite{julve3,yo2} as useful testbeds for HD gravity.

\bigskip

HD fermionic field theories have also been considered in the
literature\footnote{Fermionic theories in which the field
equations are differential equations of order greater than one are
referred as HD fremionic theories.}, for example in the context of
the effective action for the trace anomaly in conformal field
theory \cite{Shapiro}, as a dynamical mechanism for fermionic mass
generation, and in the frame of Faddeev-Popov\footnote{Although
the Faddeev-Popov compensating fields for HD-gravity are not
femionic fields (their spins are 0 or 1) they are anticommuting
ones. In this sense they are different from the usual bosonic
fields treated in \cite{julve3,yo2}.} compensating Lagrangian for
HD gravity \cite{yo1}. Therefore it would be very useful to
generalize all the work done in the bosonic case to cover the
presence of fermionic fields. This is one of the goals of the
present paper. Another is to provide a general framework for the
differential order reduction methods developed for HD theories in
\cite{ferraris} and treated in \cite{julve1}-\cite{julve3}. The
starting point in the study of any field theory, including HD
ones, is the characterization of the propagating DOF. There are
several ways to do this. The first, and standard, follows a
detailed analysis of the Ostrogradski phase space \cite{julve3},
\cite{Ostro} in order to characterize the reduced phase space
\cite{henn}, i.e. the subspace of the phase space where the
physical degrees of freedom (DOF) reside. However when the theory
is linear there exists a second and highly useful shortcut
provided by the use of the covariant symplectic techniques of
Witten and Crnkovi\'c \cite{witten,CR}. The idea of these methods
relies in the construction, directly through the action, of a
symplectic form on the space of solutions to the field equations
(covariant phase space) and use it to generate conserved
quantities (energy, angular momentum and so on) that characterize
the propagating DOF. The covariant symplectic techniques have
proved to be an essential tool in the classification of free
theories and also in the identification of the propagating DOF
\cite{Fer&Yo}. We will see that they also allow us to complete,
with \emph{all} generality, the order reduction in HD theories.

\bigskip

When we are dealing with HD relativistic field theories, the story
does not end once the DOF have been identified. The reason is that
HD theories can be usually reinterpreted in terms of lower
derivative (LD) ones that propagate (in the free limit) according
to standard Lagrangians. The machinery necessary for this
reinterpretation combines the use of the covariant Legendre
transform \cite{ferraris} and a subsequent diagonalization in the
fields \cite{julve1}. This procedure has been developed in various
examples: diff-invariant HD gravity \cite{julve1}, gauge fixed HD
electromagnetism \cite{julve2}, gauge-fixed gravity \cite{yo1},
and HD scalars \cite{julve3}. However, there are at least two
issues left aside in these works. The first one is the extension
of the above results to fermions, covering the general derivative
case\footnote{Within the frame of HD scalar theories and covariant
Legendre transform, reference \cite{julve3} fails to cover the
diagonalization of the general derivative situation. The problem
has been solved in \cite{yo2} avoiding the covariant Legendre
transform  by using Lagrange multipliers and symplectic covariant
techniques.}. The second and more interesting is to find a
unifying point of view that encompasses all the previous works.
The motivation of the present paper is to clarify these two
points.

\bigskip

The paper is organized as follows. After this introduction,
section \ref{A Review} is devoted  to review the use of the
covariant symplectic techniques for Dirac fields and to fix some
conventions. Section \ref{simple} deals with a simple HD fermionic
field theory in order to make it simpler to understand the general
$N$-derivative fermionic theory --under the hypothesis of
non-degenerate masses-- treated in section \ref{derivative}. We
end the paper with several comments and our conclusions (section
\ref{Conclusions}). Some details of the computations and the
general conventions are left to the appendices \ref{Conventions}
and \ref{Some Remarks}. Finally appendix \ref{Underlying}
indicates how the differential order reduction method proposed in
section \ref{derivative} can be extended to cover a wide class of
HD field theories.

%%%%%%%%%%%%%%%%%%%%%%%%%%%%%%%%%%%%%%%%%%%%%%%%%%%%%%%%%%%%%%%%%%%%%%%%%%%%%%%%%%%%%%%%%%%%%%%%%%%%%%%%%%%%%%%%%%%%%%%%%
%%%                                        A REVIEW OF THE DIRAC LAGRANGIAN                                           %%%
%%%%%%%%%%%%%%%%%%%%%%%%%%%%%%%%%%%%%%%%%%%%%%%%%%%%%%%%%%%%%%%%%%%%%%%%%%%%%%%%%%%%%%%%%%%%%%%%%%%%%%%%%%%%%%%%%%%%%%%%%

\section{\label{A Review}A Review of the Dirac Lagrangian}

Let us start with a quick review of the standard Dirac Lagrangian
in order to fix conventions and notations. As is well known,  the
free propagation of spin $1/2$ and mass $m$ modes can be described
through the Dirac Lagrangian ${\cal L}^{\s{D}}_m$, that can be
written in two equivalent forms
\begin{eqnarray*}
{\cal
L}^{\s{D}}_m&=&\frac{1}{2}\left[\overline{\psi}(i\overrightarrow{\dirac}-m)\psi
-\overline{\psi}(i\overleftarrow{\dirac}+m)\psi\right]=\overline{\psi}
(i\overrightarrow{\dirac}-m)
\psi-\partial_\mu\left(\frac{\overline{\psi}i\gamma^\mu\psi}{2}\right)\,.
\end{eqnarray*}
where $m$ is a mass parameter, $\psi$ is a Dirac spinor,
$\dirac:=\gamma^\mu\partial_\mu$, $\gamma^\mu$ are the Dirac
matrices, $\overline{\psi}:=\psi^\dagger \gamma^0$, and $\dagger$
denotes complex conjugation and transposition  (see appendix
\ref{Conventions} for a resume of the conventions). The first and
more symmetric  expression is appropriate for analytical purposes
and will be used in the following sections when we introduce the
differential reduction order methods. The second is suitable to
define the propagator. The Euler-Lagrange equations associated to
the Dirac Lagrangian are
\begin{eqnarray}
(i\dirac-m)\psi=0\,,\label{DiracEquation}
\end{eqnarray}
and the oriented $\overline{\psi}$-$\psi$ fermionic propagator is
\begin{eqnarray*}
\Delta^{\s{D}}_m:=\frac{1}{i\dirac-m}=-\frac{i\dirac+m}{\square+m^2}\,.
\end{eqnarray*}
The dynamics is governed by the field equations
(\ref{DiracEquation}) that define the Dirac covariant phase space
${\cal S}^{\s{D}}_m$. To parametrize ${\cal S}^{\s{D}}_m$ we solve
the field equations using spatial Fourier transform and the
notation introduced in appendix \ref{Conventions}. It is
straightforward \cite{Z} to prove that  $\psi\in{\cal
S}^{\s{D}}_m$ if and only if
\begin{eqnarray}
\psi(t,\vec{x})=\sum_{\alpha=1,2}
\int_{\mathbb{R}^3}\frac{d^3\vec{k}}{(2\pi)^3}\frac{m}{k^0} \left[
a^\alpha(k)u^\alpha(k)e^{-ik x}+ b^\alpha(k)v^\alpha(k)e^{ik
x}\right]\,,\label{DiracSol}
\end{eqnarray}
where  $u^\alpha(k)$, $v^\alpha(k)$ are basic spinors, $k
x:=k^0t-\vec{k}\cdot\vec{x}$, $k^0:=\sqrt{\vec{k}^2+m^2}\,$, and
the functional parameters that distinguish one solution from
another are encoded on the $a^\alpha(k)$ and $b^\alpha(k)$ fields.

\bigskip

Let us show how the symplectic covariant techniques \cite{witten}
induces a symplectic form $\Omega^{\s{D}}_m$ in the space ${\cal
S}^{\s{D}}_m$. First define the 2-form
\begin{eqnarray*}
\Omega^{\s{D}}=\int_\Sigma \omega^\mu_{\s{D}} d\sigma_\mu\,,
\end{eqnarray*}
where $d\sigma_\mu$ is the measure on a space-like hypersurface
$\Sigma$,
\begin{eqnarray*}
\omega_{\s{D}}^\mu&:=& \dd \frac{\partial {\cal
L}^{\s{D}}_m}{\partial \psi_\mu}\w \dd
\psi-\dd\overline{\psi}\w\dd \frac{\partial {\cal
L}^{\s{D}}_m}{\partial \overline{\psi}_\mu} =
i\dd\overline{\psi}\w \gamma^\mu\dd\psi\,,
\end{eqnarray*}
and $\dd$, $\wedge \kern-.75em \wedge$ are, respectively, the
exterior derivative and the wedge product in the linear manifold
${\cal S}^{\s{D}}_m$.

The density  $\omega_{\s{D}}^\mu$ is divergence-free when
restricted to ${\cal S}^{\s{D}}_m$, in other words
$\partial_\mu\omega_{\s{D}}^\mu=0$ modulo field equations. Then
the restriction $\Omega^{\s{D}}_m$ of $\Omega^{\s{D}}$ on ${\cal
S}^{\s{D}}_m$ is a well defined 2-form on this functional space,
i.e. is time-independent, and in the parametrization
(\ref{DiracSol}) can be written as\footnote{$z^*$ denotes the
complex conjugate of $z$.}
\begin{eqnarray}
\Omega^{\s{D}}_m&=&i\sum_{\alpha=1,2}\int_{\mathbb{R}^3}\frac{d^3\vec{k}}{(2\pi)^3}\frac{m}{k^0}\left[
\dd a^{\alpha*}(\vec{k})\w\dd a^\alpha(\vec{k})+ \dd
b^\alpha(\vec{k})\w\dd b^{\alpha*}(\vec{k}) \right]\,.
\label{SympDirac}
\end{eqnarray}
The symplectic form (\ref{SympDirac}) determines the propagating
DOF of the Dirac theory --the canonical pairs
$a^{\alpha*}(\vec{k})$-$a^{\alpha}(\vec{k})$ and
$b^{\alpha*}(\vec{k})$-$b^{\alpha}(\vec{k})$-- and can also be
used to derive constants of motion if one considers the following
version of the Noether theorem. When the symplectic form $\Omega$
is invariant under a group of transformations and we take a vector
$V$ tangent to an orbit of this group it is straightforward to
prove \cite{CR}, that locally $i_V\Omega=\dd H$, where $i_V\Omega$
denotes the contraction of $V$ and $\Omega$. The quantity $H$ is
the generator of the symmetry transformation corresponding to $V$.
If the action is Poincar\'e invariant we obtain in this way the
energy-momentum and the angular momentum densities (with the right
symmetries in their tensor indices) by computing $i_V\Omega$ for
vectors $V$ describing translations and Lorentz transformations
and writing the result as $\dd H$. Noticing that under a
translation of parameter $\tau^\mu$ we have $i_{V_T}\dd
a^\alpha(\vec{k})=i\tau^\mu k_\mu a^\alpha(\vec{k})$ and
$i_{V_T}\dd b^\alpha(\vec{k})=i\tau^\mu k_\mu b^\alpha(\vec{k})$,
the  energy of a solution is given by
\begin{eqnarray*}
H^{\s{D}}_m&=&\sum_{\alpha=1,2}\int_{\mathbb{R}^3}\frac{d^3\vec{k}}{(2\pi)^3}\frac{m}{k^0
}\,k^0\left[
a^{\alpha*}(\vec{k})a^\alpha(\vec{k})-b^\alpha(\vec{k})b^{\alpha*}(\vec{k})
\right]\,.
\end{eqnarray*}
It is worthwhile to point out that the minus sign between the $a$
and the $b$ fields forces us to take them as \emph{anti-commuting}
variables to ensure de positivity of the energy.

%%%%%%%%%%%%%%%%%%%%%%%%%%%%%%%%%%%%%%%%%%%%%%%%%%%%%%%%%%%%%%%%%%%%%%%%%%%%%%%%%%%%%%%%%%%%%%%%%%%%%%%%%%%%%%%%%%%%%%%%%
%%%                                         A SIMPLE HD FERMIONIC THEORY                                              %%%
%%%%%%%%%%%%%%%%%%%%%%%%%%%%%%%%%%%%%%%%%%%%%%%%%%%%%%%%%%%%%%%%%%%%%%%%%%%%%%%%%%%%%%%%%%%%%%%%%%%%%%%%%%%%%%%%%%%%%%%%%

\section{A \label{simple}simple HD fermionic theory}

Once we feel comfortable with the notation, lets us consider the
following  simple HD fermionic Lagrangian
\begin{eqnarray*}
{\cal L}^{\s{(2)}}_{m_1m_2}=\overline{\psi}(i
\overrightarrow{\dirac}-m_1)(i\overrightarrow{\dirac}-m_2)\psi\,,
\end{eqnarray*}
with real mass parameters\footnote{The problems associated with
the presence of degenerate and/or complex ``masses'' are detailed
in reference \cite{yo2}.} $m_1<m_2$. It is straightforward to
write down the HD-field equations for this model
\begin{eqnarray}
(i\dirac-m_1)(i\dirac-m_2)\psi=0\,, \label{EL2}
\end{eqnarray}
whose space of solutions, ${\cal S}^{\s{(2)}}_{m_1m_2}$, can be
parametrized in terms of a sum of Dirac fields in the form
\begin{eqnarray}
\psi=\psi_1+\psi_2\,,\label{Sum}
\end{eqnarray}
where $ (i\dirac-m_l)\psi_l=0$, for $l=1,2$. Explicitly, in the
notation given in Section \ref{A Review},
\begin{eqnarray}
\psi_l(t,\vec{x})=\sum_{\alpha=1,2}
\int_{\mathbb{R}^3}\frac{d^3\vec{k}}{(2\pi)^3}\frac{m_l}{k^0_l}
\left[ a_l^\alpha(k)u_l^\alpha(k)e^{-ik_l x}+
b_l^\alpha(k)v_l^\alpha(k)e^{ik_l x}\right]\,,\label{Param}
\end{eqnarray}
where, as in previous section,  $u^\alpha_l(k)$, $v^\alpha_l(k)$
are basic spinors defined in appendix \ref{Conventions}, $k_l
x=k_l^0t-\vec{k}\cdot\vec{x}$, $k^0_l:=\sqrt{\vec{k}^2+m_l^2}\,$,
and the parameters that distinguish one solution from another are
encoded on the $a^\alpha_l(k)$ and $b^\alpha_l(k)$ fields. Also
the HD propagator can be easily found as a sum of Dirac
propagators
\begin{eqnarray*}
\Delta^{\s{(2)}}_{m_1m_2}=\frac{1}{(i\dirac - m_1)(i\dirac -
m_2)}=\frac{1}{m_2-m_1}\left(\Delta_{m_2}^{\s{D}}-\Delta^{\s{D}}_{m_1}\right)\,.
\end{eqnarray*}
The physical interpretation is clear. The theory describes two LD
fermionic DOF, a physical one (positive contribution to the
energy) with mass $m_2$ and a Weyl ghost (negative contribution to
the energy) with mass $m_1$.

\bigskip

We can learn more about this theory by looking it through the
glass provided by the covariant symplectic techniques. The
extension of the covariant symplectic techniques to the HD field
theories was done in \cite{Aldaya}. Following \cite{Aldaya} the HD
Lagrangian has an associated 2-form
\begin{eqnarray*}
\Omega^{\s{(2)}}=\int_\Sigma \omega^\mu_{\s{(2)}} d\sigma_\mu\,,
\end{eqnarray*}
where
\begin{eqnarray*}
\omega_{\s{(2)}}^\mu&:=&\dd \left(\frac{\partial{\cal
L}^{\s{(2)}}_{m_1m_2}}{\partial \psi_\mu
}-\partial_\nu\frac{\partial{\cal L }^{\s{(2)}}_{m_1m_2}}{\partial
\psi_{\mu\nu}}\right)\w \dd \psi+\dd\frac{\partial{\cal
L}^{\s{(2)}}_{m_1m_2}}{\partial \psi_{\mu\nu}}\w \dd \psi_\nu\\
&=&-i(m_1+m_2)\dd\overline{\psi}\w
\gamma^\mu\dd\psi+(\partial^\mu\dd\overline{\psi})\w\dd\psi
-\dd\overline{\psi}\w\partial^\mu\dd\psi\,.
\end{eqnarray*}
It is easy to show that $\omega_{\s{(2)}}^\mu$ is real and, modulo
field equations, $\partial_\mu\omega^\mu_{\s{(2)}}=0 $. Then we
can compute the symplectic form, $\Omega^{\s{(2)}}_{m_1m_2}$, over
the space of propagating DOF, ${\cal S}^{\s{(2)}}_{m_1m_2}$,
making use of the parametrization given by equations
(\ref{Sum})-(\ref{Param}). This leads to a well defined 2-form on
${\cal S}^{\s{(2)}}_{m_1m_2}$, namely
\begin{eqnarray*}
\Omega^{\s{(2)}}_{m_1m_2}=
(m_2-m_1)\left(\Omega^{\s{D}}_{m_2}-\Omega^{\s{D}}_{m_1}\right)\,,
\end{eqnarray*}
and Noether's theorem gives us the energy
\begin{eqnarray*}
H^{\s{(2)}}_{m_1m_2}=(m_2-m_1)\left(H^{\s{D}}_{m_2}-H^{\s{D}}_{m_1}\right)\,.
\end{eqnarray*}
This supports the intuitive interpretation of ${\cal
L}^{\s{(2)}}_{m_1m_2}$ given by the propagator decomposition as a
theory describing two Dirac fields. In fact the simplectic form
confirms that this is the right interpretation and also, through
the Noether theorem, that the field $\psi_1$ is a Weyl ghost. It
contributes to the energy with a wrong sign, destroying the
semi-boundedness  of the Hamiltonian and consequently the
unitarity  of the quantum formulation of the theory.

\bigskip

Finally let us show how it is possible to find a covariant
Legendre transform connecting the HD theory with a LD one where
the usual Dirac fields are explicit. To this end we define the
first order differential operator $\D:=i\dirac$, that satisfies
$\int_{\s{\mathbb{R}^4}}
\overline{\psi}_1(D\psi_2)=\int_{\s{\mathbb{R}^4}}\overline{(D\psi_1)}\psi_2$,
and rewrite the Lagrangian in terms of this object. Namely, modulo
total derivatives,
\begin{eqnarray*}
{\cal
L}^{\s{(2)}}_{m_1m_2}(\psi,\overline{\psi},\D\psi,\overline{\D\psi}):=\overline{\D\psi}\D\psi
-\frac{m_1+m_2}{2}\left(\overline{\D\psi}\psi+\overline{\psi}\D\psi\right)
+m_1m_2\overline{\psi}\psi\,.
\end{eqnarray*}
Then we introduce a generalized Legendre transform \cite{ferraris}
with respect to the momenta generated by the
$D$-operator\footnote{We must treat a field and its conjugate as
independent variables in the Legendre transform in order to allow
the presence of grassmanian fields \cite{yo1}. Then, the
derivatives with respect $D\psi$ and $\overline{D\psi}$ are
respectively left and right derivatives.}
\begin{eqnarray}
\pi:=\frac{\partial{\cal L}^{\s{(2)}}_{m_1m_2}}{\partial
\overline{\D\psi}}=\D\psi-\frac{m_1+m_2}{2}\psi\quad;\quad
\overline{\pi}:=\frac{\partial{\cal
L}^{\s{(2)}}_{m_1m_2}}{\partial
\D\psi}=\overline{\D\psi}-\frac{m_1+m_2}{2}\overline{\psi}\,.
\label{Leg2}
\end{eqnarray}
This transformation is not singular and it permits the inversion
of the derivative of $\psi$ as a function $D\psi=\nu(\pi,\psi)$
given by
\begin{eqnarray*}
\nu(\pi,\psi):=\pi+\frac{m_1+m_2}{2}\psi\quad;\quad
\overline{\nu(\pi,\psi)}:=\overline{\pi}+\frac{m_1+m_2}{2}\overline{\psi}\,.
\end{eqnarray*}
The Legendre transform  (\ref{Leg2}) has an associated Hamiltonian
\begin{eqnarray*}
{\cal
H}^{\s{(2)}}_{m_1m_2}(\psi,\overline{\psi},\pi,\overline{\pi})&:=&
\overline{\pi} \nu(\pi,\psi) +\overline{\nu(\pi,\psi)}\pi-{\cal
L}^{\s{(2)}}_{m_1m_2}(\psi,\overline{\psi},\nu(\pi,\psi),\overline{\nu(\pi,\psi)})
\\
&=&\overline{\pi}\pi+\frac{m_1+m_2}{2}\left(\overline{\psi}\pi+\overline{\pi}\psi\right)+
\left(\frac{m_1-m_2}{2}\right)^2\overline{\psi}\psi\,,
\end{eqnarray*}
and the HD dynamics (\ref{EL2}) can be described now by means of
the Euler-Lagrange equations derived from the Helmholtz LD
Lagrangian
\begin{eqnarray*}
{\cal
L}_H^{\s{(2)}}(\psi,\overline{\psi},\pi,\overline{\pi};\D\psi,\overline{\D\psi}
):=\overline{\pi}\D\psi+\overline{\D\psi}\pi-{\cal
H}^{\s{(2)}}_{m_1m_2}(\psi,\overline{\psi},\pi,\overline{\pi})\,.
\end{eqnarray*}
Although ${\cal L}_H^{\s{(2)}}$ is a LD Lagrangian classically
equivalent to ${\cal L}^{\s{(2)}}_{m_1m_2}$, it does not
explicitly  exhibit the propagating DOF. However the
diagonalization can be carried out taking into account the special
structure of the covariant phase space. Defining the new fields
$\psi_1$ and $\psi_2$ in the form
\begin{eqnarray}
\psi&=:&\psi_1+\psi_2\,, \label{dg1}\\
\pi&=:&\frac{m_1-m_2}{2}\left(\psi_1+\psi_2\right)\,,\label{dg2}
\end{eqnarray}
the Helmholtz Lagrangian decouples the propagating modes
\begin{eqnarray*}
{\cal L}_H^{\s{(2)}}=(m_2-m_1)\left({\cal L}^{\s{D}}_{m_2}-{\cal
L}^{\s{D}}_{m_1}\right) \,.
\end{eqnarray*}
The relative minus sign shows the presence of a Weyl ghost.

\bigskip

The diagonalization (\ref{dg1})-(\ref{dg2}) can be found by means
of a very natural reasoning. Over the solution space, where the
DOF reside, we can write $\psi=\psi_1+\psi_2$ with
$(i\dirac-m_l)\psi_l=0$. Then, on this space
\begin{eqnarray*}
\psi=\psi_1+\psi_2\quad;\quad
\pi=\D\psi-\frac{m_1+m_2}{2}\psi=\frac{m_1-m_2}{2}(\psi_1-\psi_2)\,.
\end{eqnarray*}
The above relations, that coincide with the proposed
diagonalization, indicates how the DOF are encoded within the
fields $\psi$ and $\pi$. Moreover, the relation
$\{\psi,\pi\}\leftrightarrow\{\psi_1,\psi_2\}$ is invertible
because of the invertibility of the Legredre transform
(\ref{Leg2}).

\bigskip

We are ready to generalize the above results, obtained in a very
simple 2-derivative framework, to the general $N$-derivative
situation.

%%%%%%%%%%%%%%%%%%%%%%%%%%%%%%%%%%%%%%%%%%%%%%%%%%%%%%%%%%%%%%%%%%%%%%%%%%%%%%%%%%%%%%%%%%%%%%%%%%%%%%%%%%%%%%%%%%%%%%%%%
%%%                                                  N-DERIVATIVE                                                     %%%
%%%%%%%%%%%%%%%%%%%%%%%%%%%%%%%%%%%%%%%%%%%%%%%%%%%%%%%%%%%%%%%%%%%%%%%%%%%%%%%%%%%%%%%%%%%%%%%%%%%%%%%%%%%%%%%%%%%%%%%%%

\section{$N$-\label{derivative}Derivative Fermionic Theory}

In this section we consider the general $N$-derivative Lagrangian
\begin{eqnarray*}
{\cal L}^{\s{(N)}}_{\bf{m}}:=\overline{\psi}\,
\prod_{l=1}^N(i\overrightarrow{\dirac}-m_l)\, \psi \, ,
\end{eqnarray*}
where $(i\dirac-m_l)$ are Dirac operators with real mass
parameters $m_l$ that have been ordered following the $l$-index
ordering, i.e. $m_l<m_{l'}$ when $l<l'$.

\bigskip

The propagating DOF described by $ {\cal L}^{\s{(N)}}_{\bf{m}}$
can be read out from the algebraic decomposition for the HD
propagator $\Delta^{\s{(N)}}_{\bf{m}}$ in terms of Dirac
propagators
\begin{eqnarray*}
\Delta^{\s{(N)}}_{\bf{m}}:=
\frac{1}{\prod_{l=1}^N(i\dirac-m_l)}=\sum_{l=1}^N\frac{\Delta_{m_l}^{\s{D}}}{\prod_{l'\neq
l }^N (m_l-m_{l'})}\,.
\end{eqnarray*}
Notice that the sign alternates in the coefficients $\prod_{l'\neq
l }^N (m_l-m_{l'})$ so the occurrence of Weyl ghosts is to be
expected. Mathematically the propagating DOF are points in the
space of solutions to the field equations that we refer in the
following as ${\cal S}^{\s{(N)}}_{\bf{m}}$. The covariant phase
space ${\cal S}^{\s{(N)}}_{\bf{m}}$ is defined by the
Euler-Lagrange equations
\begin{eqnarray*}
\prod_{l=1}^N(i\dirac-m_l)\, \psi=0\,.
\end{eqnarray*}
These equations can be solved by means of a linear combination of
Dirac fields in the form
\begin{eqnarray*}
\psi=\sum_{l=1}^N\psi_l\,,
\end{eqnarray*}
where
\begin{eqnarray}
(i\dirac-m_l)\psi_l=0\,.\label{ELN}
\end{eqnarray}
As in previous sections, the standard parametrization for the
$\psi_l$ fields satisfying  equation (\ref{ELN}), that provides us
with a explicit parametrization of ${\cal S}^{\s{(N)}}_{\bf{m}}$,
is
\begin{eqnarray}
\psi_l(t,\vec{x})=\sum_{\alpha=1,2}
\int_{\mathbb{R}^3}\frac{d^3\vec{k}}{(2\pi)^3}\frac{m_l}{k^0_l}
\left[ a_l^{\alpha}(k)u_l^{\alpha}(k)e^{-ik_l x}+
b_l^\alpha(k)v_l^\alpha(k)e^{ik_l x}\right]\,. \label{param}
\end{eqnarray}

\bigskip

The decomposition of the HD propagator as a sum of Dirac
propagators, with alternating sign coefficients, and the
decomposition of the space of solutions as a direct sum of Dirac
spaces indicates that the theory represents the propagation of
$N$-Dirac fields, some of them physical and some of them Weyl
ghosts. This is the case, and it is possible to give a more
precise proof of this fact. The Lagrangian ${\cal
L}^{\s{(N)}}_{\bf{m}}$ induces a symplectic form in the space of
spinor fields by means of
\begin{eqnarray*}
\Omega^{\s{(N)}}=\int_\Sigma \omega_{\s{(N)}}^\mu d\sigma_\mu\,,
\end{eqnarray*}
where
\begin{eqnarray}
\omega_{\s{(N)}}^\mu&:=&\dd \left(\frac{\partial{\cal
L}^{\s{(N)}}_{\bf{m}}}{\partial \psi_\mu
}-\partial_{\mu_1}\frac{\partial{\cal L
}^{\s{(N)}}_{\bf{m}}}{\partial
\psi_{\mu\mu_1}}+\cdots+(-1)^{N-1}\partial_{\mu_1}\cdots\partial_{\mu_{N-1}}\frac{\partial{\cal
L }^{\s{(N)}}_{\bf{m}}}{\partial
\psi_{\mu\mu_1\cdots\mu_{N-1}}}\right)\w \dd
\psi\label{Density}\\
& +&\dd\left(\frac{\partial{\cal L}^{\s{(N)}}_{\bf{m}}}{\partial
\psi_{\mu\nu_1}}-\partial_{\mu_1}\frac{\partial{\cal L
}^{\s{(N)}}_{\bf{m}}}{\partial
\psi_{\mu\nu_1\mu_1}}+\cdots+(-1)^{N-2}\partial_{\mu_1}\cdots\partial_{\mu_{N-2}}\frac{\partial{\cal
L }^{\s{(N)}}_{\bf{m}}}{\partial
\psi_{\mu\nu_1\mu_1\cdots\mu_{N-2}}}\right)\w \dd \psi_{\nu_1}\nonumber\\
& +&\cdots+\,\dd\frac{\partial{\cal
L}^{\s{(N)}}_{\bf{m}}}{\partial \psi_{\mu\nu_1\cdots\nu_{N-1}}}\w
\dd \psi_{\nu_1\cdots\nu_{N-1}}\nonumber
\end{eqnarray}
This is so because over the ${\cal S}^{\s{(N)}}_{\bf{m}}$ space
the 2-form density $\omega_{\s{(N)}}^\mu$ satisfies
$\partial_\mu\omega_{\s{(N)}}^\mu=0$. Hence, the restriction of
$\Omega^{\s{(N)}}$ to ${\cal S}^{\s{(N)}}_{\bf{m}}$ --that we
refer as $\Omega^{\s{(N)}}_{\bf{m}}$-- is a well defined 2-form on
this functional space, i.e. is time-independent. In fact, it is
straightforward, but highly tedious (see appendix \ref{Some
Remarks}),  to compute this restriction in the parametrization
given by (\ref{ELN}) to obtain
\begin{eqnarray}
\Omega^{\s{(N)}}_{\bf{m}}&=&
\sum_{l=1}^N\frac{\Omega^{\s{D}}_{m_l}}{\prod^N_{l\neq
l'}(m_l-m_{l'})}\,. \label{SymplHD}
\end{eqnarray}
Consequently the DOF of the theory are a sum of Dirac DOF. Finally
the energy, computed through Noether theorem, takes the form
\begin{eqnarray*}
H^{\s(N)}_{\bf{m}}&=&\sum_{l=1}^N\frac{H^{\s{D}}_{m_l}}{\prod^n_{l\neq
l'}(m_l-m_{l'})}\,.
\end{eqnarray*}
Because of the sign alternates in the coefficients $\prod^N_{l\neq
l'}(m_l-m_{l'})$, namely $sg\left(\prod^N_{l\neq
l'}(m_l-m_{l'})\right)=(-1)^{N+l}$, some of the fields are
physical (positive contribution to the energy) and  some are Weyl
ghosts (negative contribution to the energy).

\bigskip

Once we have identified the propagating DOF --and due to the
special form of the propagator, the covariant phase space, the
symplectic form, and energy that can be obtained as linear
combinations of Dirac objects-- it is plausible to presume the
existence of a mapping that transforms the original HD theory into
a sum of LD Dirac theories. To this end, as we did in section
\ref{simple}, it is convenient to follow a series of preliminary
steps. First define the differential operator $\D:=i\dirac$ and
then expand the differential kernel $\prod_{l=1}^N(\D-m_l)$
appearing in ${\cal L}^{\s{(N)}}_{\bf{m}}$ in the form
\begin{eqnarray*}
\prod_{l=1}^N(\D-m_l)\equiv \sum_{l=0}^N c_l \D^{N-l}\,,
\end{eqnarray*}
where
\begin{eqnarray*}
c_0:=1\quad;\quad c_l:=(-1)^l\sum_{a_1<\cdots<a_l} m_{a_1}\cdots
m_{a_l}\,,\quad l=1,\dots, N.
\end{eqnarray*}
A last technical remark is still in order. The heavy algebra
involved in the following forces us to treat  the $N$ odd and $N$
even cases separately. We work out in detail the $N=2n$ one and
refer to the bosonic framework considered in \cite{julve3} and
\cite{tesis} to understand certain peculiarities present for
$N=2n-1$.\footnote{In the $N$-odd case the definition of the
highest momentum yields a constraint --that can be easily taking
into account-- while the field derivative is worked out from the
next momentum definition. Once the constraint is solved, the odd
case is exactly analogue to the even one.}

\bigskip

Modulo total derivatives the HD Lagrangian ${\cal
L}^{\s{(2n)}}_{\bf{m}}$ can be rewritten in the more convenient
form
\begin{eqnarray*}
{\cal L}^{\s{(2n)}}=\sum_{l=0}^n
c_{2l}\overline{\D^{n-l}\psi}\D^{n-l}\psi+\sum_{l=1}^n
\frac{c_{2l-1}}{2}\left[\overline{\D^{n-l}\psi}\D^{n+1-l}\psi+\overline{\D^{n+1-l}\psi}\D^{n-l}\psi
\right]\,.
\end{eqnarray*}
This expression suggests the introduction of the Ostrogradski-like
variables
\begin{eqnarray}
\chi_r:=\D^{r-1}\psi\quad;\quad\overline{\chi_r}:=\overline{\D^{r-1}\psi}\quad,
\quad r=1,\dots,n\,, \label{variables}
\end{eqnarray}
and their corresponding momenta by means of a covariant Legendre
transform
\begin{eqnarray}
&&\pi_n:=\frac{\partial {\cal
L}}{\partial\overline{\D^n\psi}}=\D\chi_n+\frac{c_1}{2}\chi_n\quad;\quad
\overline{\pi}_n:=\frac{\partial {\cal L}}{\partial
\D^n\psi}=\overline{\D\chi}_n+\frac{c_1}{2}\overline{\chi}_n\,.
\label{momentos}
\\\nonumber
&& \pi_r:=\frac{\partial {\cal
L}}{\partial\overline{\D^r\psi}}+\D\pi_{r+1}\quad;\quad\quad\quad\quad\,\,
\overline{\pi}_r:=\frac{\partial {\cal L}}{\partial
\D^r\psi}+\overline{\D\pi}_{r+1}\,,\quad r=1,\cdots, n-1.
\end{eqnarray}
It is easy to show that  this Legendre transform is non-singular
due to the invertibility of the highest derivatives $D\chi_n$ and
$\overline{D\chi_n}$ in terms of the $\chi$ and $\pi$ variables,
namely
$$
\nu_n(\chi,\pi):=D\chi_n=\pi_n-\frac{c_1}{2}\chi_n\quad;\quad
\overline{\nu_n(\chi,\pi)}:=\overline{D\chi_n}=\overline{\pi_n}-\frac{c_1}{2}\overline{\chi_n}\,.
$$
Thus it is possible to find an expression for the Lagrangian in
terms of this new variables
\begin{eqnarray*}
{\cal
L}^{\s{(2n)}}_{\bf{m}}=\overline{\pi}_n\pi_n-\frac{c_1}{2}\overline{\chi}_n\chi_n
+
\sum_{r=0}^{n-1}c_{2r}\overline{\chi}_{n-r}\chi_{n-r}+\sum_{r=0}^{n-2}\frac{c_{2r-1}}{2}\left[
\overline{\chi}_{n-r}\chi_{n-1-r}+\overline{\chi}_{n-1-r}\chi_{n-r}
\right]\,.
\end{eqnarray*}
The Ostrogradski-like Hamiltonian associated with the Legendre
transform is
\begin{eqnarray*}
{\cal H}^{\s{(2n)}}_{\bf{m}}&:=&
\overline{\pi_n}\nu_n(\chi,\pi)+\overline{\nu_n(\chi,\pi)}\pi_n+\sum_{r=1}^{n-1}\overline{\pi_r}
\chi_{r+1}+\sum_{r=1}^{n-1}\overline{\chi_{r+1}}\pi_r-{\cal
L}^{\s{(N)}}_{\bf{m}}(\chi,\overline{\chi};\nu_n(\chi,\pi),\overline{\nu_n(\chi,\pi)} )\\
&=& \overline{\pi}_n\pi_n-\frac{c_1}{2}
\left[\overline{\pi}_n\chi_n+\overline{\chi}_n\pi_n\right]+\frac{c_1^2}{2}\overline{\chi}_n\chi_n+\\
&+&\sum_{r=1}^{n-1}\left[\overline{\pi}_r\chi_{r+1}+\overline{\chi}_{r+1}\pi_r\right]
-\sum_{r=0}^{n-1}c_{2r}\overline{\chi}_{n-r}\chi_{n-r}-\sum_{r=0}^{n-2}\frac{c_{2r-1}}{2}\left[
\overline{\chi}_{n-r}\chi_{n-1-r}+\overline{\chi}_{n-1-r}\chi_{n-r}
\right]\,.
\end{eqnarray*}
Finally the Helmholtz Lagrangian --the LD Lagrangian, clasically
equivalent to ${\cal L}^{\s{(2n)}}_{\bf{m}}$, generated by the
Legendre transform-- is
\begin{eqnarray*}
{\cal
L}_H^{\s{(2n)}}:=\sum_{r=1}^n\left[\overline{\pi}_r\D\chi_r+\overline{\D\chi}_r\pi_r\right]
-{\cal H}^{\s{(2n)}}_{\bf{m}}\,.
\end{eqnarray*}
As usual in the covariant Legrendre procedure \cite{julve3,yo2},
the expression for ${\cal L}_H^{\s{(2n})}$ does not exhibit the
propagating DOF. However, its straightforward to find a new set of
variables in terms of which the propagation is explicit.
Specifically it suffices to introduce the new set of
$\psi_l$-fields\footnote{At this point the $\psi_l$ fields are
simply a new set of variables. We follow the same notation used to
denote the Dirac fields in (\ref{param}). We will show that these
new fields diagonalize the Helmholtz Lagrangian in terms of the
Dirac ones.} through
\begin{eqnarray}
\chi_r&:=&\sum_{l=1}^{2n}m_l^{r-1}\psi_l\,,\label{D1}\\
\overline{\chi}_r&:=&\sum_{l=1}^{2n}m_l^{r-1}\overline{\psi}_l\,,\\
\pi_r&:=&\sum_{l=1}^{2n}\left(\sum_{k=0}^{2(n-r)}c_km^{2n-r-k}_l
+\frac{c_{2(n-r)+1}}{2}m^{r-1}_l\right)\psi_l
\,,\\
\overline{\pi}_r&:=&\sum_{l=1}^{2n}\left(\sum_{k=0}^{2(n-r)}c_km^{2n-r-k}_l
+\frac{c_{2(n-r)+1}}{2}m^{r-1}_l\right)
\overline{\psi}_l\,.\label{D2}
\end{eqnarray}
In this variables, and modulo total derivatives, the Helmholtz
Lagrangian becomes diagonal
\begin{eqnarray*}
{\cal L}^{\s{(2n)}}_H=\sum_{l=1}^{2n}\prod_{l'\neq
l}^{2n}(m_l-m_{l'}) {\cal L}^{\s{D}}_{m_l}\,.
\end{eqnarray*}
The key observation to find the diagonalization
(\ref{D1})-(\ref{D2}) relies in the following simple fact. The
Ostrogradski variables $\chi_r$ and momenta $\pi_r$ can be
expressed by means of a one-to-one map in terms of the original HD
variable $\psi$ and its derivatives $\D^k\psi$, namely
\begin{eqnarray}
\chi_r&=&\D^{r-1}\psi\,,\label{R1}\\
\pi_r&=&\sum_{k=0}^{2(n-r)}c_k\D^{2n-r-k}\psi
+\frac{c_{2(n-r)+1}}{2}\D^{r-1}\psi\,.\label{R2}
\end{eqnarray}
When we restrict the relations (\ref{R1})-(\ref{R2}) to the space
of solutions parametrized by (\ref{param}) --that is, over the
propagating DOF space-- we reobtain the one-to-one linear
relations (\ref{D1})-(\ref{D2}) between the $\{\chi_r,\pi_r\}$ and
$\{\psi_l\}$ variables. This is the reason why this linear
redefinition diagonalizes ${\cal L}^{2n}_H$.

\bigskip

All the preceding results can be generalized to cover a wide class
of linear theories including those considered in previous works
\cite{julve2} and \cite{yo2}. In particular the procedure followed
here resolves the deficiencies inherent to more primitive
approaches \cite{julve3}. Specifically it leads us to the
diagonalization for the $N$-derivative theory within the framework
of the covariant Legendre mapping. We summarize in appendix
\ref{Underlying} the essential steps and requirements that allow
us,  following the lines of this section, to transform a HD linear
theory into a LD one where the propagating DOF are explicit.

%%%%%%%%%%%%%%%%%%%%%%%%%%%%%%%%%%%%%%%%%%%%%%%%%%%%%%%%%%%%%%%%%%%%%%%%%%%%%%%%%%%%%%%%%%%%%%%%%%%%%%%%%%%%%%%%%%%%%%%%%
%%%                                              CONCLUSIONS AND COMMENTS                                             %%%
%%%%%%%%%%%%%%%%%%%%%%%%%%%%%%%%%%%%%%%%%%%%%%%%%%%%%%%%%%%%%%%%%%%%%%%%%%%%%%%%%%%%%%%%%%%%%%%%%%%%%%%%%%%%%%%%%%%%%%%%%

\section{\label{Conclusions}Conclusions and Comments}

We have proved the equivalence of HD fermionic field theories and
a LD counterpart where the DOF are explicit  by means of a
covariant Legendre transform \cite{ferraris} and a subsequent
diagonalization. The previous attempts to solve this kind of
problems, that considered only scalar field theories
\cite{julve3},  failed to find the general diagonalization to an
arbitrary differential order due to the heavy algebra involved. A
way out of this  was given in \cite{yo2}, in a more general
framework than \cite{julve3}  --and also dealing with bosonic
field theories--, but the solution proposed there abandons the use
of the Legendre transform in favor of the Lagrange multipliers
method. The use of the covariant symplectic techniques in
combination with the covariant Legendre transform avoids the
algebraic problems of \cite{julve3} permitting us to find an
explicit formula for the DOF diagonalization at every differential
order and to generalize the previous results to the fermionic
case.

\bigskip

The approach that we follow here is exportable to more general
theories --classical mechanics, HD gravity, and so on-- as we
schematically show in appendix \ref{Underlying}. It is also
important to realize that the presence of gauge symmetries in the
theory  does not change  the results in any way. When dealing with
HD theories written in terms of differential forms, such as those
considered in \cite{yo2}, the role of the $D$-operator
($D=i\dirac$ in the present work) can be simulated by an operator
constructed by means of the exterior differential $d$ and its dual
$\delta$, for example $D=\delta d$. Even considering the possible
existence of gauge symmetries ($D d\Lambda=0$, for any field
$\Lambda$) the analogue of the Legendre mapping
(\ref{variables})-(\ref{momentos}) is still nonsingular and
therefore the reduction order procedure works exactly as in
section \ref{derivative}.

\bigskip

Another remarkable fact is that, in spite of the fact that it is
always possible to reduce the differential order by means of a
covariant Legendre mapping, the diagonalization of the resulting
LD theory in terms of a sum of standard theories it is not always
possible \cite{yo2} (but when possible, the method presented here
allow us to find it). To fix ideas, let us return to the fermionic
case where the more general HD, Lorentz invariant, Lagrangian for
a spinor $\psi$ without internal indices takes the form
$$
{\cal
L}^{\s(n)}=\overline{\psi}\left(D^n+c_1D^{n-1}+\cdots+c_{n-1}D+c_n
\right)\psi\,,
$$
where $c_i$ are real parameters with the appropriate dimensions.
As is well known, the polynomial
$D^n+c_1D^{n-1}+\cdots+c_{n-1}D+c_n$ can be factorized in terms of
its roots. In section \ref{derivative} we have make the assumption
 that all the roots are different\footnote{We have
assumed also their positivity. This is a necessary requirement if
we want these parameters to be physical masses, but it can be
relaxed for the diagonalization purposes.}. This is a necessary
hypothesis in order to succeed in the diagonalization process.
However, if we consider a HD ``mass-degenerate'' model, the
diagonalization cannot be carried out. For example, starting with
$$
{\cal L}^{\s{(2)}}_{mm}=\overline{\psi}(D-m)^2\psi\,,
$$
the Legendre procedure presented in section \ref{derivative} leads
us to the LD Hemholtz Lagrangian
$$
{\cal
L}_H^{\s{(2)}}=\left(\begin{array}{cc}\overline{\psi}&\overline{\pi}
\end{array}\right)\left(\begin{array}{cc} 0&D-m\\ D-m&-1
\end{array}\right)\left(\begin{array}{c}\psi\\\pi\end{array}\right)\,.
$$
The propagator associated with ${\cal L}_H^{\s{(2)}}$ is
$$
\Delta_{(\psi\,\pi)}=\frac{1}{(D-m)^2}\left(\begin{array}{cc} 1&D-m\\
D-m&0
\end{array}\right)\,.
$$
As it is clear, this propagator cannot be diagonalized by means of
a linear redefinition on the fields $(\psi,\pi)$ not involving
differential operators. This is so because $1/(D-m)^2$ is not a
linear combination of Dirac propagators and then it is not
possible to find a $c$-number matrix $Q$ such that $Q^\dagger
\Delta_{(\psi\,\pi)} Q$ becomes diagonal
 with Dirac propagators on its diagonal elements. The same
conclusion can be reached with an elementary analysis of the
covariant phase space.

\bigskip

Finally, we want to point out that the diagonalization process of
the  LD equivalent theory, obtained by Legendre transform,
strongly relies on the decomposition of the HD solution space as a
direct sum of LD ones. This decomposition is tied to the
decomposition of the HD propagator as a sum of simpler pieces or,
at least, on its relationships with simpler propagators. Then, in
order to make practical use of the differential order reductions
presented in the paper, the knowledge of the properties of the
Green functions of HD differential operators is needed. Some
results in this direction can be found in \cite{Avr2} where, by
means of heat kernel methods, the Green functions of a wide class
of products of second order differential operators have been
studied and their relationship with LD counterparts pointed out.
These results are a key ingredient if one wants to find the
LD-diagonalized equivalent of such a HD theory by means of a
covariant Legendre transform.

\appendix
%%%%%%%%%%%%%%%%%%%%%%%%%%%%%%%%%%%%%%%%%%%%%%%%%%%%%%%%%%%%%%%%%%%%%%%%%%%%%%%%%%%%%%%%%%%%%%%%%%%%%%%%%%%%%%%%%%%%%%%%%
%%%                                                  APPENDIX A: CONVENTIONS                                          %%%
%%%%%%%%%%%%%%%%%%%%%%%%%%%%%%%%%%%%%%%%%%%%%%%%%%%%%%%%%%%%%%%%%%%%%%%%%%%%%%%%%%%%%%%%%%%%%%%%%%%%%%%%%%%%%%%%%%%%%%%%%

\section{\label{Conventions} Conventions}
We use the Minkowski metric $\eta_{\mu\nu}=diag(+1,-1,-1,-1)$ to
lower and raise space-time indices and use the Einstein convention
for summation over repeated indices. The derivatives with respect
to coordinates are sometimes abbreviated as
$$
\partial_{\mu_1\cdots
\mu_k}:=\frac{\partial^k\hspace{.75cm}}{\partial_{\mu_1}\cdots\partial_{\mu_k}}\,,
$$
and
$\square:=\partial_\mu\partial^\mu=\partial_0^2-\vec{\partial}^2$
is the d'alambertian operator.
\bigskip

For any $X^{\mu_1\cdots\mu_k}$, $X^{\{\mu_1\cdots\mu_k\}}$
represents its index symmetrization, namely
$$
X^{\{\mu_1\cdots\mu_k\}}:=\frac{1}{k!}\sum_{\pi\in
\Pi^{\s{(k)}}}X^{\mu_{\pi_1}\cdots\mu_{\pi_k}} \,,
$$
where $\Pi^{\s{(k)}}$ denotes the permutation group of
$k$-elements.

\bigskip

The Dirac $\gamma$ matrices satisfy
$$
\gamma^{\{\mu_1}\gamma^{\mu_2\}}=\eta^{\mu_1\mu_2}
$$
with $\gamma^0$ hermitian and $\gamma^i$ antihermitian. As usual
$\kk:=k_\mu\gamma^\mu$.

\bigskip

Dirac spinors $u^\alpha(k)$ and $v^\alpha(k)$ are a basis of
solutions to the Dirac equations
$$
\left(\kk-m\right)u^\alpha(k)=0\quad;\quad
\left(\kk+m\right)v^\alpha(k)=0\,.
$$
They depend implicitly on the mass $m$ and explicitly on the
on-shell momentum $k$ with $k^0:=\sqrt{\vec{k}^2+m^2}$, and the
$\alpha$ index that labels the polarizations. In this basis, and
making use of the spatial Fourier transform, any spinor field can
be expressed as
$$
\psi(t,\vec{x})=\sum_{\alpha=1,2}\int_{\mathbb{R}^3}\frac{d^3\vec{k}}{(2\pi)^3}\frac{m}{k^0}\left[
a^\alpha(k,t)u^\alpha(k)e^{i\vec{k}\cdot\vec{x}}+
b^\alpha(k,t)v^\alpha(k)e^{-i\vec{k}\cdot\vec{x}}\right]\,.
$$
The functions $a^\alpha(k,t)$ and $b^\alpha(k,t)$ take the special
form $a^\alpha(k,t)=a^\alpha(k)e^{-ik^0t}$ and
$b^\alpha(k,t)=b^\alpha(k)e^{ik^0t}$ when considering spinor
fields that satisfy the Dirac equation $(i\dirac -m)\psi=0$.

\bigskip

 The basic spinors $u$ and $v$ satisfies the following
normalization properties
\begin{eqnarray*}
&&\bar{u}^\alpha(k)u^\beta(k)=\delta_{\alpha\beta}\quad;\quad\bar{v}^\alpha(k)v^\beta(k)=-\delta_{\alpha\beta}
\quad;\quad\bar{v}^\alpha(k)u^\beta(k)=\bar{u}^\alpha(k)v^\beta(k)=0\quad.
\\
&&u^{\alpha\dagger}(k)u^\beta(k)=\delta_{\alpha\beta}\frac{\sqrt{\vec{k}^2+m^2}}{m}\quad;\quad
v^{\alpha\dagger}(k)v^\beta(k)=\delta_{\alpha\beta}\frac{\sqrt{\vec{k}^2+m^2}}{m}\quad.
\\
&&
v^{\alpha\dagger}(k^0,-\vec{k})u^\beta(k^0,\vec{k})=u^{\alpha\dagger}(k^0,-\vec{k})v^\beta(k^0,\vec{k})=0\quad,
\end{eqnarray*}
where the conjugate spinors $\overline{u}$ are defined as
$\overline{u}:=u^\dagger \gamma^0$, and $\dagger$ denotes
transposition and complex conjugation.

%%%%%%%%%%%%%%%%%%%%%%%%%%%%%%%%%%%%%%%%%%%%%%%%%%%%%%%%%%%%%%%%%%%%%%%%%%%%%%%%%%%%%%%%%%%%%%%%%%%%%%%%%%%%%%%%%%%%%%%%%
%%%                             SOME REMARKS ON THE COMPUTATION OF THE SYMPLECTIC FORM                                %%%
%%%%%%%%%%%%%%%%%%%%%%%%%%%%%%%%%%%%%%%%%%%%%%%%%%%%%%%%%%%%%%%%%%%%%%%%%%%%%%%%%%%%%%%%%%%%%%%%%%%%%%%%%%%%%%%%%%%%%%%%%

\section{\label{Some Remarks}Some Remarks on the Computation of the Symplectic Form}
We summarize here the main steps involved in the calculation of
the symplectic form (\ref{SymplHD}) of Section \ref{derivative}.
The symplectic form is given \cite{Aldaya} through space
integration of the density
\begin{eqnarray*}
\omega_{\s{(N)}}^\mu&:=&\dd \left(\frac{\partial{\cal
L}^{\s{(N)}}_{\bf{m}}}{\partial \psi_\mu
}-\partial_{\mu_1}\frac{\partial{\cal L
}^{\s{(N)}}_{\bf{m}}}{\partial
\psi_{\mu\mu_1}}+\cdots+(-1)^{N-1}\partial_{\mu_1}\cdots\partial_{\mu_{N-1}}\frac{\partial{\cal
L }^{\s{(N)}}_{\bf{m}}}{\partial
\psi_{\mu\mu_1\cdots\mu_{N-1}}}\right)\w \dd
\psi\\
& &+\,\dd\left(\frac{\partial{\cal L}^{\s{(N)}}_{\bf{m}}}{\partial
\psi_{\mu\nu_1}}-\partial_{\mu_1}\frac{\partial{\cal L
}^{\s{(N)}}_{\bf{m}}}{\partial
\psi_{\mu\nu_1\mu_1}}+\cdots+(-1)^{N-2}\partial_{\mu_1}\cdots\partial_{\mu_{N-2}}\frac{\partial{\cal
L }^{\s{(N)}}_{\bf{m}}}{\partial
\psi_{\mu\nu_1\mu_1\cdots\mu_{N-2}}}\right)\w \dd \psi_{\nu_1}\nonumber\\
& &+\cdots+\,\dd\frac{\partial{\cal
L}^{\s{(N)}}_{\bf{m}}}{\partial \psi_{\mu\nu_1\cdots\nu_{N-1}}}\w
\dd \psi_{\nu_1\cdots\nu_{N-1}}\nonumber\,.
\end{eqnarray*}
In order to compute the derivatives of the Lagrangian that appear
in the definition of $\omega^\mu_{\s{(N)}}$ is convenient to
rewrite
\begin{eqnarray*}
{\cal L}^{\s{(N)}}_{\bf
m}(\psi,\psi_{\mu_1},\dots,\psi_{\mu_1\cdots\mu_N}):=\overline{\psi}\prod^N_{l=1}(i\dirac-m_l)\psi=\sum_{k=0}^N
c_{N-k}i^k\overline{\psi}\dirac^k\psi\,,
\end{eqnarray*}
where the constants $c_k$ are defined in terms of mass products
$$
c_0:=1\quad;\quad c_k:=(-1)^k\sum_{a_1<\cdots<a_k} m_{a_1}\cdots
m_{a_k}\,.
$$
Thus, in this notation,
\begin{eqnarray}
\frac{\partial {\cal L}^{\s{(N)}}_{\bf{m}}}{\partial
\psi_{\mu_1\cdots
\mu_k}}=c_{N-k}i^k\overline{\psi}\gamma^{\{\mu_1}\cdots\gamma^{\mu_k\}}\,,\label{derivada}
\end{eqnarray}
where the symmetrization of the Dirac gamma products can be
expressed in terms of the inverse of the Minkowskian metric
$\eta^{\mu\nu}$ in one of the following forms
\begin{eqnarray}
&&\gamma^{\{\mu_1}\cdots\gamma^{\mu_{2k}\}}=\eta^{\{\mu_1\mu_2}\cdots
\eta^{\mu_{2k-1}\mu_{2k}\}}\quad({\rm even})\label{simetrizacion}\,,\\
&&\gamma^{\{\mu_1}\cdots\gamma^{\mu_{2k-1}\}}=\gamma^{\{\mu_1}\eta^{\mu_2\mu_3}\cdots
\eta^{\mu_{2k-2}\mu_{2k-1}\}}\quad({\rm
odd})\,,\quad\quad\nonumber k\in\mathbb{N}.
\end{eqnarray}
Plugging (\ref{simetrizacion}) into (\ref{derivada}) and taking
care of the combinatorics, we found that the terms of the density
$\omega^\mu_{\s{(N)}}$ belong to one of the following four
categories
\begin{eqnarray*}
&\bf{1.}&\quad
\partial_{\mu_1\cdots\mu_{2k-1}}\frac{\partial {\cal
L}^{\s{(N)}}_{\bf{m}}}{\partial \psi_{\mu\mu_1\cdots
\mu_{2k-1}}}=(-1)^kc_{N-2k}\square^{k-1}\partial^\mu\overline{\psi}\,.\\
&{\bf 2.}&\quad
\partial_{\mu_1\cdots\mu_{2k}}\frac{\partial {\cal
L}^{\s{(N)}}_{\bf{m}}}{\partial \psi_{\mu\mu_1\cdots
\mu_{2k}}}=\frac{(-1)^kic_{N-2k-1}}{2k+1}\left(\square^{k}\partial^\mu\overline{\psi}\gamma^\mu
+2k\square^{k-1}\partial^\mu\partial_\sigma\overline{\psi}\gamma^\sigma
\right)\,.\\
&{\bf 3.}&\quad
\partial_{\mu_1\cdots\mu_{2k}}\frac{\partial {\cal
L}^{\s{(N)}}_{\bf{m}}}{\partial \psi_{\mu\nu\mu_1\cdots
\mu_{2k}}}=\frac{(-1)^{k+1}c_{N-2k-2}}{2k+1}\left(\square^k\eta^{\mu\nu}\overline{\psi}+
2k\square^{k-1}\partial^\mu\partial^\nu\overline{\psi}\right)\,.\\
&{\bf 4.}&\quad
\partial_{\mu_1\cdots\mu_{2k-1}}\frac{\partial {\cal
L}^{\s{(N)}}_{\bf{m}}}{\partial \psi_{\mu\nu\mu_1\cdots
\mu_{2k-1}}}=\frac{(-1)^{k}ic_{N-2k-1}}{2k+1}\left(
\partial^\mu\square^{k-1}\overline{\psi}\gamma^\nu
+
\partial^\nu\square^{k-1}\overline{\psi}\gamma^\mu\right.\\
&&\hspace{4.2cm}\left.+\,\,\eta^{\mu\nu}\partial_\sigma\square^{k-1}\overline{\psi}\gamma^\sigma
+2(k-1)
\partial^\mu\partial^\nu\partial_\sigma\square^{k-2}\overline{\psi}\gamma^\sigma\right)\,.
\end{eqnarray*}
Now making use of the parametrization (\ref{param}), that is
$\psi=\sum_{l=0}^N \psi_l$, where
\begin{eqnarray*}
\psi_l(t,\vec{x})=\sum_{\alpha=1,2}
\int_{\mathbb{R}^3}\frac{d^3\vec{k}}{(2\pi)^3}\frac{m_l}{k^0_l}
\left[ a_l^{\alpha}(k)u_l^{\alpha}(k)e^{-ik_l x}+
b_l^\alpha(k)v_l^\alpha(k)e^{ik_l x}\right]\,,\quad
k_l^0=\sqrt{\vec{k}^2+m^2}\,,
\end{eqnarray*}
remembering that the constants $c_k$ are defined in terms of the
masses $m_l$ and using the properties of the spinors $u_l$ and
$v_l$ detailed in Appendix \ref{Conventions} it is
straightforward, but slightly tedious, to derive equation
(\ref{SymplHD}) for the symplectic form.

%%%%%%%%%%%%%%%%%%%%%%%%%%%%%%%%%%%%%%%%%%%%%%%%%%%%%%%%%%%%%%%%%%%%%%%%%%%%%%%%%%%%%%%%%%%%%%%%%%%%%%%%%%%%%%%%%%%%%%%%%
%%%                                              THE UNDERLYING IDEA                                                  %%%
%%%%%%%%%%%%%%%%%%%%%%%%%%%%%%%%%%%%%%%%%%%%%%%%%%%%%%%%%%%%%%%%%%%%%%%%%%%%%%%%%%%%%%%%%%%%%%%%%%%%%%%%%%%%%%%%%%%%%%%%%

\section{\label{Underlying}The Underlying Idea}

The procedure developed in main body of the paper for spinorial
fields can be straightforwardly extended to cover all linear
models for which the field equations can be derived from a
variational principle of the form
\begin{eqnarray}
S_{\s{H\!D}}[\phi]=\langle \phi\,|\,(D+M_1)\cdots
(D+M_{\s{N}})\phi\rangle\,,\label{HD}
\end{eqnarray}
where $M_i$ are ``mass''  parameters (usually $M_i=-m_i$ when
dealing with fermions, $M_i=m^2_i$ when dealing with bosons),
ordered by their subscript  $M_1<M_2<\cdots<M_N$,
$\langle\cdot|\cdot\rangle$ is a pseudo-scalar product, and $D$ a
differential operator symmetric under $\langle\cdot|\cdot\rangle$,
that is
\begin{eqnarray*}
\langle \phi_1\,|\,D\phi_2\rangle =\langle
D\phi_1\,|\,\phi_2\rangle\,.
\end{eqnarray*}
Generically, in the fermionic case $D$ will be a first order
differential operator such as $D=i\dirac$ and in the bosonic case
a second order one such as $D=\delta d$ for differential forms or
$D=\square\left(\frac{1}{2}P^{(2)}-P^{(s)}\right)$ for HD-gravity
\cite{yo2}.

\bigskip

  Under this hypothesis, and through the lines presented in
Section \ref{derivative}, it is a trivial task to define a
correspondence between $S_{\s{H\!D}}[\phi]$ and the action
\begin{eqnarray}
\sum_{a=1}^N\prod_{b\neq a}^N (M_b-M_a)S_a[\phi_a ]\,,\label{LD}
\end{eqnarray}
where
\begin{eqnarray*}
S_a[\phi_a]:=\langle\phi_a\,|\, (D+M_a)\phi_a\rangle\,.
\end{eqnarray*}
This can be done by following the same steps as in the spinorial
case. First define a Legendre transformation of the form
(\ref{variables})-(\ref{momentos}) and then a linear redefinition
in the form of (\ref{D1})-(\ref{D2}) to select the propagating
DOF. Formally, the same formulas are valid in the general case
with the obvious identifications $\psi\leftrightsquigarrow \phi$,
$\overline{\psi}\leftrightsquigarrow \langle \phi|$, and
$m_l\leftrightsquigarrow -M_l$.

%%%%%%%%%%%%%%%%%%%%%%%%%%%%%%%%%%%%%%%%%%%%%%%%%%%%%%%%%%%%%%%%%%%%%%%%%%%%%%%%%%%%%%%%%%%%%%%%%%%%%%%%%%%%%%%%%%%%%%%%%
%%%                                                 ACKNOWLEDGEMENTS                                                  %%%
%%%%%%%%%%%%%%%%%%%%%%%%%%%%%%%%%%%%%%%%%%%%%%%%%%%%%%%%%%%%%%%%%%%%%%%%%%%%%%%%%%%%%%%%%%%%%%%%%%%%%%%%%%%%%%%%%%%%%%%%%

\begin{acknowledgments}

The author wishes to thank F. Barbero,  J. Julve, and F. J. de
Urries for several interesting discussions. This work was
supported by a Spanish Ministry of Education and Culture
fellowship co-financed by the European Social Fund.

\end{acknowledgments}

%%%%%%%%%%%%%%%%%%%%%%%%%%%%%%%%%%%%%%%%%%%%%%%%%%%%%%%%%%%%%%%%%%%%%%%%%%%%%%%%%%%%%%%%%%%%%%%%%%%%%%%%%%%%%%%%%%%%%%%%%
%%%                                               BIBLIOGRHAPHY                                                       %%%
%%%%%%%%%%%%%%%%%%%%%%%%%%%%%%%%%%%%%%%%%%%%%%%%%%%%%%%%%%%%%%%%%%%%%%%%%%%%%%%%%%%%%%%%%%%%%%%%%%%%%%%%%%%%%%%%%%%%%%%%%

\end{document}